\documentclass[12pt]{iopart}

\usepackage{iopams}
\usepackage{dsfont}
\usepackage{amsthm}
\theoremstyle{definition}
\newtheorem{Def}{Definition}[section]

\theoremstyle{plain}
\newtheorem{Lem}[Def]{Lemma}
\newtheorem{Prop}[Def]{Proposition}
\newtheorem{Thm}[Def]{Theorem}

\theoremstyle{remark}

\newtheorem*{Pf}{Proof}
\newtheorem*{Thm*}{}

\newcommand{\dd}{\mathrm{d}}
\newcommand{\ee}{\mathrm{e}}

\newcommand{\rh}{r_\mathrm{h}}
\newcommand{\Ie}{I_\varepsilon}
\newcommand{\We}{\stackrel{\begin{minipage}{5mm}
                 \scriptsize$\eps\quad$\vspace*{-1.6mm}\end{minipage}}
                {W^{1,2}}}
\newcommand{\Wb}[2]{\stackrel{\begin{minipage}{5mm}
                 \scriptsize$\eps\quad$\vspace{-2.3mm}\end{minipage}}
                {W^{1,2}_{\,#1,#2}}}                
\newcommand{\eps}{\varepsilon}
\newcommand{\Xe}{X^\varepsilon}
\newcommand{\R}{\mathds R}
\newcommand{\norm}[1]{\|#1\|}
\newcommand{\artanh}{\mathrm{artanh}}
\newcommand{\expan}{\theta_{(l)}}

\begin{document}

\title[A universal inequality for axisymmetric and stationary black holes]
 {A universal inequality between angular momentum and horizon area
 for axisymmetric and stationary black holes
 with surrounding matter}

\author{J\"org Hennig, Marcus Ansorg and Carla Cederbaum}

\address{Max Planck Institute for Gravitational Physics,\\
Am M\"uhlenberg 1, D-14476 Golm, Germany}
\eads{\mailto{pjh@aei.mpg.de}, \mailto{mans@aei.mpg.de}
      and \mailto{Carla.Cederbaum@aei.mpg.de}}

\begin{abstract}
We prove that for sub-extremal axisymmetric and stationary black
holes with arbitrary surrounding matter the inequality
$8\pi|J|<A$ holds, where
$J$ is the angular momentum and $A$ the horizon area of the black hole.

\end{abstract}

\pacs{04.70.Bw, 04.40.-b, 04.20.Cv\hfill preprint number: AEI-2008-034}

%\submitto{\CQG}
%\maketitle

%%%%%%%%%%%%%%%%%%%%%%%%%%%%%%%%%%%%%%%%%%%%%%%%%%%%%%%%%%%%%%%%%%%%%%%%%
\section{Introduction}

A well-known property of the Kerr solution, describing a single rotating
black hole in vacuum, is given by
\begin{equation}\label{KN}
|p_J| \le 1 \quad\textrm{with}\quad
p_J:=\frac{8\pi J}{A},
\end{equation}
where $J$ and  $A$ denote the angular momentum and  the horizon area of
the black hole respectively. Equality in \eref{KN} holds if and only 
if the Kerr black hole is \emph{extreme}.
As was shown in \cite{Ansorg}, the
equation $|p_J|=1$ is even true more generally for axisymmetric and
stationary black holes with surrounding matter in the
degenerate limit (i.e. for vanishing surface gravity $\kappa$).
Moreover, it was also conjectured in \cite{Ansorg} that
$|p_J| \le 1$ still holds if the black hole is
surrounded by matter. In this paper we prove this 
conjecture\footnote{Note that in \cite{Ansorg} a more general conjecture, 
incorporating the black hole's electric charge $Q$, was formulated. 
Here we prove this conjecture for the pure Einstein field, 
i.e. for $Q=0$, and vanishing cosmological constant $\Lambda=0$.
(It should be noted that, for $\Lambda\neq 0$, the inequality 
$|p_J|\leq 1$ can be violated. 
An example is the Kerr-(A)dS family
of black holes, see \cite{Booth}.)}.

We start by requiring that a physically relevant non-degenerate black
hole be characterized  
through the existence of trapped surfaces (i.e. surfaces with negative
expansion $\expan$ 
of outgoing null geodesics) in an interior vicinity of the event
horizon. That is, in the  
terminology of \cite{Booth},
we concentrate on {\em sub-extremal} black holes. In the following
we show that such surfaces cannot exist for $|p_J|\ge 1$, provided that
an appropriate functional $I$  
(to be defined below) cannot fall below 1. In turn, this can be proved
by means of  
methods from the calculus of variations.

%%%%%%%%%%%%%%%%%%%%%%%%%%%%%%%%%%%%%%%%%%%%%%%%%%%%%%%%%%%%%%%%%%%%%%%%%
\section{Coordinate systems and Einstein equations}

We introduce suitable coordinates and metric functions by adopting our
notation from \cite{Ansorg}, which is based on \cite{Bardeen}.
In spherical coordinates $(r,\theta,\varphi,t)$, an exterior vacuum 
vicinity of the event horizon\footnote{For a stationary space-time, 
the immediate vicinity of a black hole event horizon must be vacuum, see
e.~g.~\cite{Bardeen}.} 
can be described by the line element
\begin{equation}\label{LE1}
  \dd s^2 = \ee^{2\mu}(\dd r^2+r^2\dd\theta^2)
            +r^2 B^2\ee^{-2\nu}\sin^2\!\theta\,
            (\dd\varphi-\omega\dd t)^2 - \ee^{2\nu}\dd t^2.
\end{equation}
The function $B$ must be
a solution of $\nabla\cdot(r\sin\theta\,\nabla B)=0$, with
$Br\sin\theta=0$ at the event  
horizon $\mathcal H$. Here $\nabla$
is the nabla operator in three-dimensional flat space.
In order to fix the coordinates we chose the particular solution
$B=1-\rh^2/r^2$, where $\rh=\textrm{constant}>0$.
In this manner we
obtain coordinates in which $\mathcal H$ is located at the coordinate sphere
$r=\rh$.
We now decompose the potential $\nu$ in the form
$\nu = u + \ln B$, thereby
%\begin{equation}
%\nu = u + \ln B
%\end{equation}
obtaining three regular metric functions $\mu$, $u$ and $\omega$,
depending on $r$ and $\theta$ only.

Following Bardeen~\cite{Bardeen}, we introduce the new metric
potentials $\hat\mu = r^2\ee^{2\mu}$, $\hat u = r^2 \ee^{-2 u}$,
%\begin{equation}
% \hat\mu = r^2\ee^{2\mu},\quad
% \hat u = r^2 \ee^{-2 u},
%\end{equation}
which are positive and regular functions of $r$ and $\cos\theta$,
as well as the new radial coordinate
$R=\frac{1}{2}\left(r+\frac{r_\mathrm{h}^2}{r}\right)$.
%\begin{equation}
% R=\frac{1}{2}\left(r+\frac{r_\mathrm{h}^2}{r}\right).
%\end{equation}
We arrive at the Boyer-Lindquist type line element
\begin{equation}\label{LE2}
 \dd s^2 = \hat\mu\left(\frac{\dd R^2}{R^2-\rh^2}+\dd\theta^2\right)
           +\hat u\sin^2\!\theta\,(\dd\varphi-\omega\dd t)^2
           -\frac{4}{\hat u}(R^2-\rh^2)\dd t^2, 
\end{equation}
which is singular at $\mathcal H$ ($R=\rh$).

In these coordinates, the vacuum Einstein equations
read as follows:\footnote{Throughout this paper we consider pure gravity, i.e. no
electromagnetic fields, as well as vanishing cosmological constant, $\Lambda=0$.}
\begin{equation}\label{EG1}
\fl (R^2-\rh^2)\tilde u_{,RR}+2R\tilde u_{,R}+\tilde u_{,\theta\theta}
 +\tilde u_{,\theta}\cot\theta 
  = 1 - \frac{\hat{u}^2}{8}\sin^2\!\theta\!
  \left(\omega_{,R}^2+\frac{\omega_{,\theta}^2}{R^2-\rh^2}\right)\!\!,\
\end{equation}
\begin{equation}\label{EG2}
\fl  (R^2-\rh^2)(\omega_{,RR}+4\omega_{,R}\tilde u_{,R})
 +\omega_{,\theta\theta}+\omega_{,\theta}(3\cot\theta+4\tilde u_{,\theta})=0,
\end{equation}
\begin{eqnarray}\label{EG3}
 \fl (R^2-\rh^2)\tilde\mu_{,RR}+R\tilde\mu_{,R}
 +\tilde\mu_{,\theta\theta}
 & = & \frac{\hat{u}^2}{16}\sin^2\!\theta
  \left(\omega_{,R}^2+\frac{\omega_{,\theta}^2}{R^2-\rh^2}\right)
   + R\tilde u_{,R}\nonumber\\
  & &  -(R^2-\rh^2)\tilde u_{,R}^2
  -\tilde u_{,\theta}(\tilde u_{,\theta}+\cot\theta),
\end{eqnarray}
where $\tilde u:=\frac{1}{2}\ln\left(\rh^{-2}\hat u\right)$ and
$\tilde\mu:=\frac{1}{2}\ln\left(\rh^{-2}\hat\mu\right)$. At the horizon, the metric potentials
obey the boundary conditions \cite{Bardeen}
\begin{equation}\label{BC}
 \mathcal H:\quad \omega=\textrm{constant}=\omega_\mathrm{h},\quad
                  \frac{2\rh}{\sqrt{\hat\mu\hat u}}
                  = \textrm{constant}=\kappa,
\end{equation}
with the horizon angular velocity $\omega_\mathrm{h}$ and the
surface gravity $\kappa$. On the horizon's north and south pole ($R=\rh$
and $\sin\theta=0$), 
the following regularity conditions hold:
\begin{equation}\label{RC}\fl
 \hat\mu(R=\rh,\theta=0)=\hat u(R=\rh,\theta=0) = \hat\mu(R=\rh,\theta=\pi)=\hat u(R=\rh,\theta=\pi) 
= \frac{2\rh}{\kappa}.
\end{equation}
%%%%%%%%%%%%%%%%%%%%%%%%%%%%%%%%%%%%%%%%%%%%%%%%%%%%%%%%%%%%%%%%%%%%%%%%%
\section{Necessary condition for the existence of trapped surfaces}

A crucial quantity for the following discussion is the expansion
$\expan=h^{ab}\nabla_a l_b$ of outgoing null rays for two-surfaces
$\mathcal S$ in an interior vicinity of the horizon, where $h$ is
the interior
metric of $\mathcal S$ and $l$ is the vector field describing
outgoing null rays. In order to analyze $\expan$ inside the black hole, we
introduce horizon-penetrating coordinates
$(R,\theta,\tilde\varphi,\tilde t\,)$,
in which the metric is \emph{regular} at the horizon $\mathcal H$:
\begin{equation}
\dd\tilde t = \dd t+\frac{T(R)}{\Delta}\dd R, \quad
\dd\tilde\varphi = \dd\varphi + \frac{\Phi(R)}{\Delta}\dd R,\quad
\Delta:=4(R^2-\rh^2).
\end{equation}
The free functions $T$ and $\Phi$ are chosen in such a way that
\begin{equation}
 a(R,\theta):=\frac{4\hat\mu\hat u-T^2}{\Delta},\quad
 b(R,\theta):=\frac{\Phi-\omega T}{\Delta}
\end{equation}
are regular at $R=\rh$. Furthermore, we require $a>0$ in order to
guarantee that $\tilde t=\textrm{constant}$ is a \emph{spacelike} surface.
 As a consequence, $T$ has to obey the conditions
\begin{equation}\label{T}
 T=2\sqrt{\hat\mu\hat u}=\frac{4\rh}{\kappa}\quad\mbox{for $R=\rh$},
 \qquad
 T>2\sqrt{\hat\mu\hat u}\quad\mbox{for $R<\rh$}.
\end{equation}
We thus obtain the regular line element\footnote{Note that for the Kerr
solution with mass $M$ and angular momentum $J$, 
we obtain Kerr type coordinates (where the slices
$\tilde t=\textrm{constant}$ are spacelike)
by choosing $T=4M(M+2R)$ and
$\Phi=\textrm{constant}=2J/M$.}
\begin{eqnarray}
 \dd s^2 & = & \left(\frac{a}{\hat u}
             +\hat u b^2\sin^2\!\theta\right)\dd R^2
             -\frac{\Delta}{\hat u} \dd{\tilde t}^{\,2}
             +\hat u\sin^2\!\theta(\dd\tilde\varphi-\omega\dd\tilde t\,)^2
             +\hat\mu\dd\theta^2
             \nonumber\\
         &   & +2\left(\frac{T}{\hat u}+\omega\hat u
             b\sin^2\!\theta\right)\dd R\dd\tilde t
             -2\hat u b\sin^2\!\theta\,\dd R\dd\tilde\varphi. 
\end{eqnarray}
In these coordinates, we calculate the expansion $\expan$ for a
surface $\mathcal S$ in a small interior
neighborhood of the horizon, described by
\begin{equation}\label{Surface_S}
 \mathcal S:\quad
 R=\rh-\varepsilon\hat r(\theta), \quad \theta\in[0,\pi], \quad
 \tilde\varphi \in [0,2\pi), \quad\tilde t = \textrm{constant},
\end{equation}
where $\varepsilon>0$, $\hat r>0$ and $\varepsilon\hat r\ll\rh$.
We obtain
\begin{equation}\label{theta}
 \expan(\theta)  = \frac{\hat T}{2\sqrt{2\hat\mu a}\sin\theta}
          \left.\left[\frac{(\hat
          r_{,\theta}\sin\theta)_{,\theta}}{\hat r\rh}
          -\frac{(\hat\mu\hat u)_{,R}}{2\hat\mu\hat
          u}\sin\theta\right]\right|_{\mathcal H}\varepsilon
          +\mathcal O(\varepsilon^2),
\end{equation}
where we have used that $T$ is of the form $T=2\sqrt{\hat\mu\hat
u}\big|_\mathcal{H}+\varepsilon\hat T$ with
$\hat T=\hat T(\theta)>0$ on
$\mathcal S$; see \eref{T}.
[Note that $\expan=0$ on the horizon ($\varepsilon=0$).]

Following Booth and Fairhurst \cite{Booth}, we study the criterion
$\delta_{\bar n}\theta_{(\bar l)}<0$  (in their notation)
for the existence of trapped surfaces,
i.e. we characterize a physically relevant, sub-extremal black hole 
through a negative variation of the expansion on the horizon 
in direction of an ingoing null field $\bar n$.
In our formulation, this is equivalent to the existence of a surface
$\mathcal S$ with negative expansion $\expan$.
%It now turns out that
We first show the following lemma.
\begin{Lem}\label{Lem1}
A necessary condition for the existence of trapped surfaces in the
interior vicinity of the event horizon of an
axisymmetric and stationary black hole is
 \begin{equation}\label{lemma}
  \int\limits_0^\pi(\hat\mu\hat
  u)_{,R}\big|_\mathcal{H}\sin\theta\,\dd\theta>0.
 \end{equation}
\end{Lem}

\begin{Pf}
Let the surface $\mathcal S$, defined in (\ref{Surface_S}) for sufficiently small $\varepsilon$,
be trapped. Then
$\expan$ is negative everywhere on $\mathcal S$, and so is the term in square brackets in
\eref{theta} for all $\theta\in[0,\pi]$. Thus, 
the integral of this term along the horizon $\mathcal H$ is
negative. Integrating by parts, and using that $\hat\mu\hat
u=\textrm{constant}>0$ on $\mathcal H$ [see \eref{BC}], yields
\begin{equation}
 \frac{1}{\rh}\int\limits_0^\pi\frac{\hat r_{,\theta}^2}{\hat
 r^2}\sin\theta\,\dd\theta
 -\frac{1}{2\hat\mu\hat u}\int\limits_0^\pi(\hat\mu\hat
 u)_{,R}\sin\theta\,\dd\theta<0.
\end{equation}
Since the first integral is non-negative, we immediately obtain
\eref{lemma}.\quad $\Box$
\end{Pf}
%%%%%%%%%%%%%%%%%%%%%%%%%%%%%%%%%%%%%%%%%%%%%%%%%%%%%%%%%%%%%%%%%%%%%%%%%
\section{Calculation of $p_J$}

Following the notation of \cite{Ansorg}, we express 
angular momentum $J$ and horizon area $A$ of the black hole by
\begin{eqnarray}
 \fl J  &=&  \frac{1}{8\pi}\oint\limits_\mathcal{H}m^{a;b}\dd S_{ab}
     = -\frac{1}{16}\int\limits_0^\pi\hat u^2\omega_{,R}\big|_\mathcal{H}
       \sin^3\!\theta\,\dd\theta
     =  -\frac{\rh^2}{4}\int\limits_{-1}^1V\ee^{2U}(1-x^2)\dd x,\\
 \fl A &=& 2\pi\int\limits_0^\pi\sqrt{\hat\mu\hat u}\big|_\mathcal{H}
     \sin\theta\,\dd\theta
   = 4\pi\sqrt{\hat\mu\hat u}\big|_\mathcal{H}=4\pi\rh^2\ee^{2U(1)},
\end{eqnarray}
where $x:=\cos\theta$, 
$U(x) := \left.\left[\frac{1}{2}\ln\left(\rh^{-2}\hat
u\right)\right]\right|_\mathcal{H}$\,, 
$V(x) := \left.\left[\frac{1}{4}\,\hat u\,\omega_{,R}\right]
\right|_\mathcal{H}$\,,
and $m^a$ is the Killing vector corresponding to axisymmetry. Note that
for this formulation we have used  
conditions \eref{BC} and \eref{RC}. As a consequence, $p_J$ takes the form
\begin{equation}\label{S}
 p_J\equiv \frac{8\pi J}{A} = -\frac{1}{2}\ee^{-2U(1)}
 \int\limits_{-1}^1 V\ee^{2U}(1-x^2)\dd x.
\end{equation}

%%%%%%%%%%%%%%%%%%%%%%%%%%%%%%%%%%%%%%%%%%%%%%%%%%%%%%%%%%%%%%%%%%%%%%%%%
\section{Reformulation in terms of a variational problem}

In order to prove the inequality in question, we show that
for $|p_J|\ge 1$ Eq.~\eref{lemma} is violated, i.e. that there are no
trapped surfaces in the interior vicinity of the horizon.

Using \eref{EG1} and \eref{EG3}, the integrand in \eref{lemma} can be
expressed 
\begin{equation}\label{muu}
 (\hat\mu\hat u)_{,R}\big|_\mathcal{H}
 = \frac{2}{\rh}\hat u^2\big|_{\theta=0}
  \left[1-\frac{1}{16}\omega_{,R}^2\hat u^2\sin^2\!\theta
  - \frac{\hat u_{,\theta}}{2\hat u}
    \left(\frac{\hat u_{,\theta}}{2\hat
    u}+2\cot\theta\right)\right]\Big|_\mathcal H.
\end{equation}
Hence, we can write \eref{lemma} in terms of
$U$, $V$ and $x$ as follows:
\begin{equation}\label{lemma1a}
 \frac{1}{2}\int\limits_{-1}^1
   \left[(V^2+U'^{\,2})(1-x^2)-2xU'\right]\dd x<1
   \quad \mbox{with}\quad U':=\frac{\dd U}{\dd x}.
\end{equation}
From \eref{S}, the violation of \eref{lemma} for $|p_J|\ge 1$ can 
thus be formulated as the following implication, to be valid for any
regular functions $U,V$ defined on $[-1,1]$ and satisfying the condition
$U(-1)=U(1)$ [which follows from (\ref{RC})]: 
\begin{equation}\label{int}\fl
\left|\; \int\limits_{-1}^1 V\ee^{2U}(1-x^2)\dd x\right| \ge 2\ee^{2U(1)}
 \quad \Rightarrow\quad
 \frac{1}{2}\int\limits_{-1}^1
   \left[(V^2+U'^{\,2})(1-x^2)-2xU'\right]\!\dd x \ge 1.\
\end{equation}
We now show that the validity of this implication holds provided that an
appropriate functional $I$ (defined below) cannot fall below 1. 

Applying the Cauchy-Schwarz inequality
to the first inequality in \eref{int} we obtain
\begin{equation}\label{CSU}\fl
 4\ee^{4U(1)} \le
 \left(\;\int\limits_{-1}^1 V\ee^{2U}(1-x^2)\dd x\right)^2 \le
 \int\limits_{-1}^1V^2(1-x^2)\dd x\int\limits_{-1}^1\ee^{4U}(1-x^2)\dd x.
\end{equation}
Given this inequality, we replace the term $\int V^2(1-x^2)\dd x$ 
in the second inequality in \eref{int} and see that 
\begin{equation}\label{functional}\fl
I[U]:=\frac{1}{2}\int\limits_{-1}^1
   \left[U'^{\,2}(x)(1-x^2)-2xU'(x)\right]\dd x
   +\frac{2\ee^{4U(1)}}
    {\int\limits_{-1}^1\ee^{4U(x)}(1-x^2)\dd x}\geq 1
\end{equation}
is as a sufficient condition
for the validity of the implication,
with the functional $I: W^{1,2}(-1,1)\to \R$ defined on the Sobolev
space $W^{1,2}(-1,1)$. 
Thus we have shown the following.
\begin{Lem}\label{Lem2}
 The inequality $|p_J|<1$ for
 any sub-extremal axisymmetric and stationary black hole with surrounding
 matter holds provided that the inequality
 \begin{equation}\label{lemma2}
  I[U] \ge 1 
 \end{equation}
is true for all $U\in W^{1,2}(-1,1)$ with $U(-1)=U(1)$.
\end{Lem}
This reformulation has led us to the following \emph{variational
problem}: Calculate the 
minimum of the functional $I$ and show that it is not
below $1$. 

%%%%%%%%%%%%%%%%%%%%%%%%%%%%%%%%%%%%%%%%%%%%%%%%%%%%%%%%%%%%%%%%%%%%%%%%%
\section{Complete solution of the variational problem}

Consider the functional
\begin{equation}\label{functional1}\fl
 I_\varepsilon[U]
  :=\frac{1}{2}\int\limits_{-1+\varepsilon}^{1-\varepsilon}
   \left[U'^{\,2}(x)(1-x^2)-2xU'(x)\right]\dd x
   +\frac{2\ee^{4U(1-\varepsilon)}}
    {\int\limits_{-1+\varepsilon}^{1-\varepsilon}
    \ee^{4U(x)}(1-x^2)\dd x}
\end{equation}
on $W^{1,2}(-1+\eps,1-\eps)$, where
$0<\varepsilon \ll 1$ is a fixed real number. We use techniques from the
calculus of variations to show that there exists a minimizer
$U_\varepsilon$ for $\Ie$ in a suitable class
 with sufficiently large value
$\Ie[U_\varepsilon]$. Following this investigation, we take the 
limit $\varepsilon\to0$ and see that the claim of lemma \ref{Lem2}
follows. 

We now show the following statements:
\begin{enumerate}

 \item The functional $\Ie$ is \emph{well-defined on the Sobolev space
       $\We:=W^{1,2}(-1+\eps,1-\eps)$} of functions $U$
       defined almost everywhere on $(-1+\eps,1-\eps)$. 
       This is due to the well-known proposition below.
       \begin{Prop}[Thm 2.2 in Buttazzo-Giaquinta-Hildebrandt
         \cite{Buttazzo}]\label{prop}
          On any bounded interval $J\subseteq\R$, $W^{1,2}(J)\hookrightarrow
          C^0(\overline{J})$ compactly. Moreover, the fundamental theorem of
          calculus holds in $\We$. 
       \end{Prop}
       Here, we use the adapted inner product
       $\int_{-1+\eps}^{1-\eps}UV\,\dd\mu
        +\int_{-1+\eps}^{1-\eps}U'V'\,\dd\mu$
       for $U,V\in\,\We$, where $\dd\mu(x):=(1-x^2)\,\dd x$,
       which is equivalent to the ordinary one
       %by continuity
       %and as $2\eps(1-\eps) \leq 1-x^2 \leq 1$ on $(-1+\eps,1-\eps)$,
       and thus makes $\We$ a Hilbert space. 

       For ease of notation set
       $\Xe:=\{U\in\,\We|\,U(1-\eps)=0\}$.
       Note that the functional $\Ie$
       is \emph{invariant under addition of constants}.
       %i.e. $\Ie[U+b]=\Ie[U]$ for all constants $b\in\R$
       %and all $U\in\,\We$.
       We will use this in order to
       restrict our attention to the Hilbert subspace $X^\eps$ on which
       $(U,V):=\int_{-1+\eps}^{1-\eps}U'V'\,\dd\mu$ is an
       equivalent inner product inducing the norm
       $\norm{U}:=\big(\int_{-1+\eps}^{1-\eps}
        U'^{\,2}\,\dd\mu\big)^{1/2}$. 

  \item $\Ie$ is \emph{bounded from below}. Using
        $0\le \left(\frac{x}{\sqrt{1-x^2}}-U'(x)\sqrt{1-x^2}\right)^2
          = \frac{x^2}{1-x^2}-2xU'(x)+U'^{\,2}(x)(1-x^2)$
        we conclude that
        $I_\varepsilon[U]\ge -\frac{1}{2}\int_{-1+\eps}^{1-\eps}
        \frac{x^2}{1-x^2}\,\dd x =:C(\eps)> -\infty$.

  \item Applying the Cauchy-Schwarz inequality to
        $\int_{-1+\varepsilon}^{1-\varepsilon} xU'(x)\dd x$,
        we obtain 
        that $\Ie[U] \ge \frac{1}{2}\norm{U}^2
        +2C(\eps)\norm{U}$ for any $U\in\Xe$
        with $C(\eps)$ as in (ii).
        Hence, for every
        $P\in\mathds R$ there exists a $Q_P\in\mathds R$ such that
        $\Ie[U] \ge P$ whenever $\norm{U}\ge Q_P$.
        This is equivalent to \emph{coercivity of the functional $\Ie$
        with respect to the weak topology on $\Xe$.}

  \item The functional $\Ie$ is {\it sequentially lower
        semi-continuous (lsc) with respect to the weak topology in $\Xe$}.
        %i.e. for each weakly converging sequence
        %$U_k\rightharpoonupU$, we have that
        %$\liminf_{k\to\infty} I_\eps(U_k)\geq
        %I_\eps(U)$.
        Recall that lower semi-continuity is additive and that the first
        terms can be dealt with by standard theory (see e.g. \cite{Yosida}).
        %As the notion of sequential lower semi-continuity is additive, we can
        %consider each term of $\Ie$ individually. The first term is just
        %one half times the square of the norm and whence lsc by standard
        %theory (cf. e.g. \cite{Alt}). The middle term can be
        %interpreted to be one half times the inner product with $x\mapsto
        %\int_{1-\eps}^x\frac{y}{1-y^2}dy\in \Xe$ and is hence lsc by
        %definition of weak convergence.
        For the last term, we use proposition
        \ref{prop} to deduce that $U_k\to U$ in
        $C^0([-1+\eps,1-\eps])$. Whence there exists a uniform bound $D>0$
        of $\{U_k\}$ so that by Lipschitz continuity of the
        exponential map on $[-4D,4D]$ with Lipschitz constant $L$, we have
        \begin{eqnarray} 
          \left|\int_{-1+\eps}^{1-\eps}\ee^{4U_k}\dd\mu
         -\int_{-1+\eps}^{1-\eps}\ee^{4U}\dd\mu\right|
         &\leq &
         4L\int_{-1+\eps}^{1-\eps}|U_k-U|\dd\mu
         \nonumber\\
         &\leq & 8L\norm{U_k-U}
         _{C^0([-1+\eps,1-\eps])}
         \stackrel{k\to\infty}{\longrightarrow}0.
       \end{eqnarray}
  The last term thus being lsc, we have shown $\Ie$ to be lsc.      
  
\end{enumerate}

We can now show \emph{existence of a global minimizer for
$\Ie$ in a suitable class}:\\  
As we have seen in (ii), $\Ie$ is bounded from below on
$\Xe$. Hence for $a\in\R$ and $c>0$
we can choose a minimizing sequence $\{U_k\}$ in the class
$\mathcal{K}^\eps_{a,c}:=\{U\in \Xe\,\vert\,
\int_{-1+\eps}^{1-\eps} \ee^{4U}\,\dd\mu=c,\ U(-1+\eps)=a\}$,
%i.e. a sequence where $\Ie[U_k]$ tends
with values tending
to the infimum $i^\eps_{a,c}$
of $\Ie$ in $\mathcal{K}^\eps_{a,c}$. By coercivity,
%[choose $P:=|i^\eps_{a,c}|+1>0$ in (iii)],
$\{U_k\}$ is bounded and we can
extract a weakly converging subsequence with limit
$U^\eps_{a,c}\in \Xe$ by Hilbert space techniques (theorem
of Eberlein-Shmulyan \cite{Yosida}).
%We denote this subsequence by $\{U_k\}$ for
%notational convenience. 

The class $\mathcal{K}^\eps_{a,c}$ is weakly sequentially closed by
proposion \ref{prop}, which can be shown as in (iv).
Whence by (iv), $U^\eps_{a,c}\in\mathcal K^\eps_{a,c}$ satisfies
$\Ie[U^\eps_{a,c}]=i^\eps_{a,c}$.
%In consequence,
%$U^\eps_{a,c}\in\mathcal{K}^\eps_{a,c}$. By sequential lower
%continuity [property (iv)], we deduce that
%$\Ie[U^\eps_{a,c}]=i^\eps_{a,c}$ so that we have identified a
%minimizer. 

Set
$\Wb{a}{0}:=\{U\in\,\We\,\vert\,U(-1+\eps)=a,\,
U(1-\eps)=0\}\subset \Xe$ for any $a\in\R$.
By the theory of Lagrange multipliers, each
minimizer of $\Ie$ in the class $\mathcal{K}^\eps_{a,c}$ is a
critical point of the functional
\begin{equation}\fl
 J_{\eps,c}:\,\Wb{a}{0}\to\R:U\mapsto
 \frac{1}{2}\int\limits_{-1+\eps}^{1-\eps}
 \left[U'^{\, 2}(x)(1-x^2)-2xU'(x)\right]\dd x
 +\frac{\lambda}{2}\left(\int_{-1+\eps}^{1-\eps}
 \ee^{4U}\dd\mu-c\right)
\end{equation}
for some $\lambda\in\R$, 
which is well-defined and sufficiently smooth by proposition
\ref{prop}. In other words, there is $\lambda:=\lambda^\eps_{a,c}\in\R$
such that $U:=U^\eps_{a,c}\in\mathcal{K}^\eps_{a,c}$ satisfies
\begin{equation}\label{EL}\fl
 \int\limits_{-1+\eps}^{1-\eps}\left[U'(x)
 \varphi'(x)(1-x^2)-x\varphi'(x)\right]\!\dd x
 +2\lambda\int\limits_{-1+\eps}^{1-\eps}\ee^{4U(x)}
 \varphi(x)(1-x^2)\dd x = 0\
\end{equation}
for all $\varphi\in\,\Wb{0}{0}$. This can be restated to say that
$U\in\,\We$ is a weak solution of

\begin{equation}\fl\label{star}
\begin{minipage}{15cm}\vspace{0.8ex}
  $-U''(x)(1-x^2)+2xU'(x)+1+2\lambda\ee^{4U(x)}(1-x^2)
  =  0\quad\forall x\in(-1+\eps,1-\eps)$,\\[-0.5ex]
 
 $U(-1+\eps) = a,\quad
 U(1-\eps)  =  0,\quad
 \int_{-1+\eps}^{1-\eps}\ee^{4U}\dd\mu = c.$
 %\vspace{0.8ex}
\end{minipage}
\end{equation}

%\begin{eqnarray}\label{star}
% \nonumber\fl
% -U''(x)(1-x^2)+2xU'(x)+1+2\lambda\ee^{4U(x)}(1-x^2)
%  =  0\quad\forall x\in(-1+\eps,1-\eps)\mbox{}\\
% \fl
% U(-1+\eps) = a,\quad
% U(1-\eps)  =  0,\quad
% \int_{-1+\eps}^{1-\eps}\ee^{4U}\dd\mu = c.
%\end{eqnarray}

Any weak solution $U\in\,\We$ of \eref{star} can be shown to be
smooth and to satisfy equation \eref{star} strongly: For all
$\varphi\in\Wb{0}{0}$, we can rewrite \eref{EL} as
\begin{equation}\label{31}
0  = \int\limits_{-1+\eps}^{1-\eps}
      \left[U'(x)(1-x^2)-x
      -2\lambda\,\int_{-1+\eps}^{x}\ee^{4U}\dd\mu\right]
      \varphi'(x)\,\dd x,
\end{equation}
where we used integration by parts and proposition
\ref{prop}. By the fundamental lemma of the calculus of variations,
there is a constant $b\in\R$ such that
\begin{equation}\label{starstar}
 U'(x)(1-x^2)-x-2\lambda\,\int_{-1+\eps}^{x}
 \ee^{4U}\dd\mu=b
 %\textrm{ a.e. on } (-1+\eps,1-\eps)
\end{equation}
holds almost everywhere on $(-1+\eps,1-\eps)$ as the integrand of
\eref{31} in square brackets is an $L^1$-function.
%As this equation can be solved for $U'(x)$, we
%deduce $U'\in C^1([-(1-\eps),1-\eps])$ again by proposition
%\ref{prop}. Inductively, a bootstrap argument guarantees
Solving for $U'$, we deduce the smoothness
of $U$
%up to the boundary
by a bootstrap argument (similar to p.~462 in
\cite{Evans}).
Differentiating Eq.~\eref{starstar},
we get strong validity of \eref{star}. In
particular, $U^\eps_{a,c}$ is a smooth classical solution of the
Euler-Lagrange equation of $J_{\eps,c}$. 

Interestingly,
there exists an integrating factor for Eq.~\eref{star} and it can be
solved explicitly:
For
\begin{equation}\fl
 F(x):=-(1-x^2)^2U'^{\,2}(x)+2x(1-x^2)U'(x)
 +\lambda\ee^{4U(x)}(1-x^2)^2-x^2
\end{equation}
we have
\begin{equation}\fl
 F'(x)=2[x-(1-x^2)U'(x)]
 \left[U''(x)(1-x^2)-2xU'(x)
 -1-2\lambda\ee^{4U(x)}(1-x^2)\right].
\end{equation}
Thus, it suffices to solve the first order equation $F=\textrm{constant}$,
which can be done with the substitution $W(x):=(1-x^2)^2\ee^{4U(x)}$.

The unique smooth solution turns out to be
\begin{equation}\label{sol}
 W(x)=\eps^2(2-\eps)^2\left(\frac{\ee^{2\alpha\,\artanh(1-\eps)}
       -\beta\ee^{-2\alpha\,\artanh(1-\eps)}}
       {\ee^{2\alpha\,\artanh\,x}-\beta\ee^{-2\alpha\,\artanh\,x}}\right)^2,
\end{equation}
where the Lagrange multiplier $\lambda$ was eliminated using the condition
$U(1-\eps)=0$.
The complex integration constants $\alpha$ and $\beta$
are implicitly given in terms of
$a$ and $c$ via $U(-1+\eps)=a$ and
$\int_{-1+\eps}^{1-\eps}\ee^{4U}\dd\mu=c$.
The corresponding value of $\Ie$ is
\begin{eqnarray}
 \fl
 \Ie[U^\eps_{a,c}] & = &
  \frac{\alpha\beta}{2}\frac{(2-\eps)^{4\alpha}-\eps^{4\alpha}}
   {(1+\beta^2)[\eps(2-\eps)]^{2\alpha}
     -\beta[(2-\eps)^{4\alpha}+\eps^{4\alpha}]}
    +1-\eps+\frac{1-\alpha^2}{2}\ln\frac{\eps}{2-\eps} \nonumber\\
   \fl
   &&
   +\frac{8\alpha[\eps(2-\eps)]^{2(\alpha-1)}}
   {[(2-\eps)^{2\alpha}-\eps^{2\alpha}]^2}
   \cdot
   \frac{(1+\beta^2)[\eps(2-\eps)]^{2\alpha}
     -\beta[(2-\eps)^{4\alpha}+\eps^{4\alpha}]}
     {(2-\eps)^{4\alpha}-\eps^{4\alpha}}.
\end{eqnarray}

We now study the limit $\eps\to 0$.
For any $U\in\,W^{1,2}(-1,1)$ set
$a_\eps:=U(-1+\eps)-U(1-\eps)$,
$c_\eps:=\int_{-1+\eps}^{1-\eps}\ee^{4U-4U(1-\eps)}
\dd\mu$,
$c:=\int_{-1}^{1}\ee^{4U}\dd\mu$.
Then
$U\big|_{(-1+\eps,1-\eps)}-U(1-\eps)\in\mathcal{K}^\eps_{a_\eps,c_\eps}$
and thus $\Ie[U]\geq i^\eps_{a_\eps,c_\eps}$. As we have seen,
there is a minimizer $U^\eps_{a_\eps,c_\eps}$ of
$\Ie$ in $\mathcal{K}^\eps_{a_\eps,c_\eps}$ and we obtain
$\Ie[U]\geq \Ie[U^\eps_{a_\eps,c_\eps}]$. As discussed
above, this minimizer is smooth and  solves the Euler-Lagrange equation
of the functional $J_{\eps,c_\eps}$;
%weakly. Moreover, the minimizer
%it is smooth up
%the boundary and solves this equation
the unique smooth solution of this
%the Euler-Lagrange
equation is given by
\eref{sol}. For $\eps\to 0$ we have $\alpha_\eps\to 1$
(otherwise $\Ie$ diverges) and $\beta_\eps\to -1$
[a consequence of $a_\eps\to 0$ by \eref{lemma2}].
Using these relations, we obtain
$\Ie[U^\eps_{a_\eps,c_\eps}]\geq C^\eps_{a_\eps}$
with $C^\eps_{a_\eps}\to1$ as $\eps\to0$.
By definition of $I$ and $\Ie$ we see that
\begin{equation}\fl
 \left|I(U)-I_\eps(U)\right| \leq 
 \frac{1}{2}\!\!\!\!\!\! \int\limits_{1-\eps \le |x|<1}\!\!\!\!\!\!
 |U'^{\,2}(x)(1-x^2)-2xU'(x)|\,\dd x
 + 2\,\left|{\frac{\ee^{4U(1)}}{\int_{-1}^1
 \ee^{4U}\dd\mu}
 -\frac{\ee^{4U(1-\eps)}}{\int_{-1+\eps}^{1-\eps}
 \ee^{4U}\dd\mu}}\right|\!,\,
\end{equation}
where the first term tends to $0$
and the denominators of the latter tend to each other as
$\eps\to0$ (the integrands are $L^1$-functions, theorem of bounded
convergence). The numerators of the latter term converge to
each other by proposition \ref{prop}. Thus,
$I[U]\geq1$.\footnote{Interestingly, equality in both inequalities of
(\ref{int}) [and hence $I(U)=1$] is achieved only for the horizon functions
$U$ and $V$ of a {\em degenerate} black hole (e.g. extreme Kerr),
cf. \cite{Ansorg}.} 

Applying Lemma~\ref{Lem2}, we have shown $|p_J|<1$ for
\emph{sub-extremal} black holes. Together with the results about
\emph{extremal} black holes \cite{Ansorg}\footnote{With the presented
{\em unique} smooth solution (\ref{sol}) of the  
differential equation (\ref{star}), the 
assumption of equatorial symmetry as well as that of the existence of a
continuous sequence in the proof presented in \cite{Ansorg} can be
abandoned, cf. equation (35) in \cite{Ansorg}.} we arrive at the
following. 

\begin{Thm}
 Consider space-times with pure gravity (no electromagnetic fields) and
 vanishing cosmological constant. Then,
 for every axisymmetric and stationary sub-extremal
 black hole with arbitrary
 surrounding matter we have that $8\pi|J| < A$.
 The equality $8\pi|J| = A$ holds if
 the black hole is degenerate (extremal).
\end{Thm}

% \section{Discussion}
% The validity of $|p_J|<1$ for non-degenerate axisymmetric and stationary
% black holes has
% consequences for other concepts in general relativity, namely for the
% definition of mass $M=\frac{J}{2}(p_J+1/p_J)$ by Christodoulou and Ruffini
% \cite{Christodoulou} and for the definition of surface gravity
% $k_r=\sqrt{\frac{p_J}{8J\big(1+p_J^2\big)}}(1-p_J^2)$ by Ashtekar
% \cite{Ashtekar}. 
% If $|p_J|$ could exceed the limit of $1$, then it would be possible to
% find different
% black hole solutions (solutions with different values of $p_J$)
% but with the same parameters $J$ and $M$,
% in contradiction
% to the \lq\lq no hair\rq\rq\ theorem.
% Furthermore, Ashtekar's surface gravity
% $k_r$ could become negative. 

%%%%%%%%%%%%%%%%%%%%%%%%%%%%%%%%%%%%%%%%%%%%%%%%%%%%%%%%%%%%%%%%%%%%%%%%%
\ack
We are indebted to Jos\'e Luis Jaramillo and
Herbert Pfister for many valuable hints and comments.
We wish to thank Paul T Allen for commenting
on the manuscript.
This work was supported by the Deutsche
Forschungsgemeinschaft (DFG) through the
Collaborative Research Centre SFB/TR7
\lq\lq Gravitational wave astronomy\rq\rq. 
%%%%%%%%%%%%%%%%%%%%%%%%%%%%%%%%%%%%%%%%%%%%%%%%%%%%%%%%%%%%%%%%%%%%%%%%%

%%%%%%%%%%%%%%%%%%%%%%%%%%%%%%%%%%%%%%%%%%%%%%%%%%%%%%%%%%%%%%%%%%%%%%%%%

\end{document}